\documentclass{PoS}

\newcommand{\vecc}[1]{\mbox{\boldmath $#1$}}

\title{Gauge-Invariant Gluon TMD and Evolution in the Coordinate Space}

\ShortTitle{Gauge-Invariant gTMD}

\author{\speaker{Igor O. Cherednikov}
\\
        EDF - Departement Fysica, Universiteit Antwerpen, Belgium\\
        E-mail: \email{igor.cherednikov@uantwerpen.be}}

\abstract{Maximally path-dependent gauge-invariant operator definition of the gluon transverse-momentum dependent pdf (gTMD) is discussed. It is argued that the evolution equations for the gTMD in the coordinate representation can be derived from the equations of motion in the generalised loop space, whose elements are the hadronic averages of the Wilson loops of entirely arbitrary shape.}

\FullConference{QCD Evolution 2015 -QCDEV2015-\\
		26 - 30 May 2015\\
		Jefferson Lab (JLAB), Newport News Virginia, USA}

\begin{document}

\section{Introduction}

Transverse-momentum dependent parton density functions (TMD pdfs in what follows) extent the idea of collinear pdfs, which accumulate mass singularities in any order of perturbative expansion, making it possible to apply the QCD factorisation approach to the inclusive hadron processes ($e^+e^-$ to hadrons, DIS) \cite{Feynman:1969ej}. In 'more differential' cross-sections, which one deals with in the semi-inclusive reactions (semi-inclusive DIS, Drell-Yan, Higgs, vector boson, heavy-flavour production) factorisation often implies {\it intrinsic} transverse-momentum dependence of non-perturbative pdfs, which must be constructed in such a way to absorb not only mass, but also rapidity singularities \cite{Collins:1981uk, Collins:1981uw, Collins:2011zzd, Collins:2003fm, Cherednikov:2007tw}. With TMD pdfs one gets access to the 
three-dimensional structure of the nucleon in the momentum space, which is actively investigated in ongoing and planned experiments at the LHC, JLab + JLab 12 GeV upgrade, RHIC, EIC etc., providing an incentive to rapid theoretical progress in the field (see, e.g., \cite{Boer:2011fh, Angeles-Martinez:2015sea} and Refs. therein).

As compared to the quark TMD case, the gauge-invariant gluon TMD exhibits much more involved structure of the gauge links (Wilson lines), so that the path-dependence brings about extra complications to the understanding of the {\it (non-)universality} property of the TMD pdfs \cite{Bomhof:2006dp, Bomhof:2007xt}. On the other hand, this path-dependence allows us to manipulate freely with the gauge links entering the operator definition of the gTMD.
In the present paper we propose a quantum-field theoretic approach to the definition of the gluon TMD (gTMD) making use of the arbitrariness of the trajectories of the Wilson lines to formulate evolution of the gTMD in terms of the equations of motion in the so-called generalised loop space, the fundamental degrees of freedom of which are the hadronic matrix elements of the arbitrary Wilson loops.

\section{Operator structure of TMD pdfs}

Following the standard approach one derives the operator definitions of the TMD pdfs starting with the 
{\it factorisation} of a given process in a convenient gauge, which naturally entails the following generic {\it gauge-dependent} correlators: 

\begin{equation}
 {\mathbf Q}_{\rm g-d} (k; P, S) 
= 
\int\! d^4 z \ {\rm e}^{- i k z} \ \langle h | \ \bar \psi (z) \psi (0) \ | h \rangle 
\end{equation}
for quarks and 
\begin{equation}
{\mathbf G}_{\rm g-d}^{\mu\nu} (k; P, S) 
=  \\
\int\! d^4 z \ {\rm e}^{- i k z} \ \langle h | \ {\cal A}^{\mu} (z) {\cal A}^{\nu} (0) \ | h \rangle 
\end{equation}
for gluons. In what follows we focus on the gluon case. 
Gauge invariance of these correlators can be achieved by inserting a Wilson line between the field operators, whose {underlying path $\gamma$} is determined by the factorisation scheme  \cite{Collins:1981uk, Collins:2011zzd, Mulders:2000sh, Belitsky:2002sm, Ji:2005nu}: 

\begin{equation}
{\mathbf G}_{\rm g-i}^{\mu \nu | \mu' \nu'} (k; P, S) 
=
\int\! d^4 z \ {\rm e}^{- i k z} \ \langle h | \ {\cal F}^{\mu \nu} (z) \ {{\cal W}_\gamma } \ {\cal F}^{\mu' \nu'} (0) \ | h \rangle , 
\label{eq:gen_corr}
\end{equation}
where the field strength is defined in the adjoint representation
\begin{equation}
{\cal F}^{\mu \nu}  =  \partial^\mu {\cal A}^\nu - \partial^\nu {\cal A}^\mu - i g [{\cal A}^\mu, {\cal A}^\nu] , \ {\cal F}_{\mu \nu} = {F}_{\mu \nu}^a T^a . 
\end{equation}

{\it Gauge-invariant} gTMD, that is a hadronic matrix element, which corresponds to the distribution of the gluons with momentum $k = (k^+,0^-,\vecc k_\perp)$ inside the hadron $h$ with momentum $P$ and spin $S$, is obtained by integrating over the minus-light-cone component of the partonic momentum and projecting onto the transverse Lorentz indices:

\begin{eqnarray}
& & {\mathbf G}^{ij} (x, \vecc k_\perp; P, S) 
\sim
\int\! d k^- \ {\mathbf G}^{+i | + j} (k; P, S) =  
\\
& & \int\! d z^- d^2 z_\perp\ {\rm e}^{- i k^+ z^- - i k_\perp z_\perp} \ \langle h | \ {\cal F}^{+i} (0^+, z^-, \vecc z_\perp) \ { {\cal W}_\gamma } \ {\cal F}^{+j} (0^+,0^-,\vecc 0_\perp) \ | h \rangle . 
\label{eq:tmd_standard}
\end{eqnarray}
Factorisation in the small-$x$ regime entails another definition of the gTMD having different structure \cite{ Dominguez:2011wm}. The latter can be related, however, to the moderate-$x$ definition (\ref{eq:tmd_standard}), see Refs. \cite{Balitsky:2015qba}. 

Let us explore a possibility of an alternative approach to the definition of the TMD pdfs. We construct first a   
generic {\it fully gauge-invariant} (by construction) and {\it maximally path-dependent} object consisting of the Wilson loops defined on the set of absolutely arbitrary trajectories $\gamma_i$: 
\begin{equation}
 {{\cal W}_{\gamma_i}} =   {\cal P}_{\gamma_i}\ {\rm exp} \left[ \oint_{\gamma_i} \ d\zeta_\mu { {\cal A}^\mu } (\zeta) \right] . 
 \end{equation}
So far we do not assume any factorisation scheme and even do not identify the processes to which those objects are supposed to be applicable. 
Given the arbitrariness of the paths, one is able to adjust them to a specific factorisation scheme by performing the   {\it evolution in the coordinate space}. Hence we end up with an {object} which can be further associated with a realistic {\it gauge-invariant} {TMD pdf}.

\section{Equations of motion in the loop space}

Shape variations (or evolution in the coordinate representation) of arbitrary Wilson loops can be consistently formulated in terms of the equations of motion in the loop space \cite{MM_WL_1, Makeenko:2002uj}. They are integral-differential equations which the elements of this space (generic Wilson loops) obey, if the underlying contours $\gamma_i$ on which the path-ordered exponentials of the gauge fields are defined experience certain variations.  The variations of the contours give rise to the variations of the exponentials themselves, the latter being described by the infinite set of the Makeenko-Migdal loop equations  \cite{MM_WL_1, Makeenko:2002uj, WL_Renorm_1, Tavares:1993pw, DeG_2014}. 
More specifically, the elements of the loop space are the vacuum matrix elements of products of the {Wilson loops}, that is 
\begin{equation}
\langle 0 |  { {\cal W}^n_{\gamma_1, ... \gamma_n} } | 0 \rangle
 =
 \langle 0 |   {\cal T}  { {\cal W}_{\gamma_1} } \cdot \cdot \cdot  {{\cal W}_{\gamma_n } } | 0  \rangle 
\end{equation}
These fundamental gauge-invariant (by construction) degrees of freedom obey the {Makeenko-Migdal} loop equations (in the large-$N_c$ limit)


\begin{equation}
 \partial^\nu \ \frac{\delta}{\delta \sigma_{\mu\nu} (x)} \ \langle 0|  { {\cal W}^1_\gamma } | 0 \rangle
 =
 N_c g^2 \ \oint_{\gamma} \ dz^\mu \ \delta^{(4)} (x - z) \langle 0 | { {\cal W}^2_{\gamma_{xz} \gamma_{zx}} } | 0 \rangle , 
\end{equation}
where the area and path differential operators are defined as follows \cite{MM_WL_1, WL_Renorm_1, Makeenko:2002uj}: 

\begin{itemize}

\item {Area derivative describes the behaviour of the Wilson loop under the infinitesimal variations of the area ${\delta \sigma_{\mu\nu}}$ at a given point}

\begin{equation}
 \frac{\delta}{\delta \sigma_{\mu\nu}(z)}  \langle 0 | {{\cal W}_\gamma} | 0 \rangle
 =
 \lim_{|\delta \sigma_{\mu\nu}(z)| \to 0} \frac{\langle 0 | {{\cal W}_{\gamma\delta\gamma_x}} |0  \rangle - \langle  0 | { {\cal W}_\gamma} | 0 \rangle}{|\delta \sigma_{\mu\nu}(z)|}
\end{equation}

\item {Path derivative deals with the infinitesimal path extensions and contractions $\delta z_\mu^{-1}\gamma\delta z_\mu$ at a given point}

\begin{equation}
 \partial_\mu  \langle 0 |  {{\cal W}_\gamma} | 0 \rangle
 =
 \lim_{|\delta z_{\mu}| \to 0} \frac{\langle 0 | { {\cal W}_{\delta z_\mu^{-1}\gamma\delta z_\mu} | 0 \rangle} - \langle 0 |  { {\cal W}_\gamma} |0 \rangle}{|\delta z_{\mu}|}
\end{equation}

\end{itemize}

These {differential operators} in the loop space determine the {\it evolution} of the Wilson loops in the coordinate representation. They can also be related to the energy/rapidity evolution of the TMDs with pure light-like Wilson lines \cite{Cherednikov:2012yd, Mertens:2013xga}.
 In other words, starting from a loop having a given shape, one can come to a loop with another shape by solving the above evolution equations. This is exactly what is needed to adjust a generic Wilson loop to some specific geometrical layout prescribed by a factorisation framework. 

\section{Stokes-Mandelstam gluon TMD}

Let us show how this strategy can be practically implemented. Making use of non-Abelian {Stokes' theorem} (see Ref. \cite{Makeenko:2002uj} and Refs. therein)
\begin{equation}
{\cal P}_\gamma\ {\rm exp} \left[ \oint_\gamma \ d\zeta_\rho {{\cal A}^\rho} (\zeta) \right]
= 
{\cal P}_\gamma {\cal P}_\sigma {\rm exp}\  \left[ \int_\sigma \ d\sigma_{\rho\rho'} (\zeta) { {\cal F}^{\rho\rho'}  } (\zeta)  \right] , 
\end{equation}
where ${\cal P}_\sigma$ stands for the area-ordering, 
and the Mandelstam formula 
\begin{equation}
\frac{\delta}{\delta \sigma_{\mu\nu} (x)} {\cal P}_\gamma\ {\rm exp} \left[ \oint_\gamma \ d\zeta_\rho {{\cal A}^\rho } (\zeta) \right]
=
{\cal P}_\gamma \  { {\cal F}^{\mu\nu}  } (x)   {\rm exp} \left[ \oint_\gamma \ d\zeta_\rho { {\cal A}^\rho } (\zeta) \right] ,
\end{equation}
we see that the generic correlation function (\ref{eq:gen_corr}) (Fourier-transformed to the coordinate space) can be represented in the following form
\begin{eqnarray}
 \tilde {\mathbf G}^{\mu \nu | \mu' \nu'} (z; P, S) 
& &  = 
\frac{\delta}{\delta \sigma_{\mu\nu} (z)} 
\frac{\delta}{\delta \sigma_{\mu'\nu'} (0)} 
\langle h |  { {\cal W}_{\gamma^{[z,0]}} }    \ | h \rangle 
 \\ 
& & = \frac{\delta}{\delta \sigma_{\mu\nu} (z)} 
\frac{\delta}{\delta \sigma_{\mu'\nu'} (0)}  \sum_X \ 
\langle h |  { {\cal W'}_{\gamma^{[z]}} } | X \rangle \langle X |   { {\cal W'}_{\gamma^{[0]} } } \ | h \rangle , 
\end{eqnarray}
where $ {\cal W}_{\gamma^{[z,0]}}$ stand for a Wilson loop with the underlying path $\gamma$ containing the points $z$ and $0$. This is the only condition for the integration trajectories, in the rest they are entirely arbitrary. 

We coin this representation the {\it Stokes-Mandelstam gTMD} definition. 
The key feature of this definition (and of the entire approach) is that one first calculates the hadronic matrix element of an arbitrary Wilson loop 
$$
\langle h |  { {\cal W}_{\gamma^{[z,0]}} }    \ | h \rangle , 
$$
choosing its shape the most convenient for the practical purposed.
In particular, in the situations where the non-Abelian {exponentiation} is applicable (see, e.g., Refs. \cite{Gatheral:1983cz})
\begin{equation}
\langle h | {\cal W}_{\gamma^{[z,0]}} | h \rangle 
=
{\rm exp} \left[ \sum a_n W^{(n)} \right] , \ W^{(n)} = {\rm hadronic \ correlators} , 
\end{equation}
one can even obtain an explicit expression for this matrix element and thereafter evaluate the area derivatives in terms of the fundamental hadronic correlation functions. The latter can be taken, for instance, from lattice simulations.
Let us illustrate the use of this framework by considering a simple Abelian example.

\subsection{Example: Abelian exponentiation}

Wilson loops in the Abelian gauge theory are known to exponentiate
\begin{eqnarray}
\langle h | {\cal W}_\gamma | h \rangle  & = & 
\langle h | {\cal P}_\gamma\ {\rm exp} \left[ \oint_\gamma \ d\zeta_\rho {  {\cal A}^\rho } (\zeta) \right] | h \rangle  \\
& = &
 {\rm exp} \left[ - \frac{g^2}{2}  \oint_\gamma \ d\zeta_\rho \oint_\gamma \ d\zeta'_{\rho'} \ { D^{\rho\rho'} (\zeta - \zeta')  } \right] , 
 \label{eq:WL_A}
\end{eqnarray}
where the basic hadronic correlator reads
\begin{equation}
{D^{\rho\rho'} (\zeta - \zeta') }
=
\langle h | {\cal P}_\gamma\  A^\rho (\zeta) A^{\rho'} (\zeta') | h \rangle .
\label{eq:bas_corr}
\end{equation}

For the sake of simplicity, let us consider a spinless hadron, which entails the following parameterisation of the basic correlator (\ref{eq:bas_corr})

\begin{equation}
D_{\rho\rho'} (z) 
 = g_{\rho\rho'}\  {D_1  (z, P) } 
+ \partial_\rho \partial_{\rho'} \  {D_2 (z, P) }
+ \{ P_\rho \partial_{\rho'} \} \  { D_3 (z, P)} + 
P_\rho P_{\rho'}  \ { D_4 (z, P) } . 
\end{equation}
Given this representation and the exponentiation, the area derivative can be evaluated straightforwardly

\begin{equation}
\frac{\delta}{\delta \sigma_{\mu\nu} (z)}  \langle h | {\cal W}_\gamma | h \rangle 
= 
- \frac{g^2}{2}  \ \left[ \frac{\delta}{\delta \sigma_{\mu\nu} (z)} \ \oint_\gamma \ d\zeta_\rho \oint_\gamma \ d\zeta'_{\rho'} \ {D_{\rho\rho'} (\zeta - \zeta')  }\right] \ 
\langle h | {\cal W}_\gamma | h \rangle 
\end{equation}

After taking the path derivative $\partial_\mu$, 
the terms containing only one derivative disappear since
\begin{equation}
\partial_\mu^z \left[ \oint_\gamma \ d\zeta'_{\mu} \partial_{\rho} {D_i (z - \zeta') } \right]
= 0 , 
\end{equation}
and the {non-vanishing terms} are

\begin{itemize}

\item the {standard `Makeenko-Migdal' term}:
\begin{equation}
\oint_\gamma \ d\zeta^\nu \ \partial^2 \ { D_1 (z, P) }
\end{equation}

\item 
the {hadron  momentum-dependent term}, obviously absent in the loop space with vacuum matrix elements

\begin{equation}
 \oint_\gamma \ d\zeta^\nu \ (P \partial)^2  \ { D_4 (z, P) }
\end{equation}

\end{itemize}

Therefore, the {\it shape evolution} equation for the hadronic Wilson loops in the Abelian gauge theory is given by

\begin{eqnarray}
& &  \partial_\mu^z \frac{\delta}{\delta \sigma_{\mu\nu} (z)}  \langle h |{ {\cal W}_\gamma } | h \rangle 
= \\
& &
- \frac{g^2}{2}  \left[ \oint_\gamma \ d\zeta^\nu \ \Big( \partial^2 \ {D_1 (z, P) } + (P \partial)^2  \ {D_4 (z, P) } \Big)  \right] \langle h | {  {\cal W}_\gamma }| h \rangle .
\label{eq:SM_MM}
\end{eqnarray}

Taking into account that for the vacuum matrix elements
$$
\partial^2 D_1 (z) = - \delta^{(4)} (z) , 
$$
and that $D_4 = 0$ in vacuum, 
one easily re-obtains the Makeenko-Migdal equation in the leading order: 
\begin{equation}
\partial_\mu^z \frac{\delta}{\delta \sigma_{\mu\nu} (z)}  \langle 0 |{ {\cal W}_\gamma } | 0 \rangle 
=
g^2 \oint_\gamma\! d\zeta^\nu \ \delta^{(4)} (z - \zeta) ,
\label{eq:MM_1}
\end{equation}
as well as the area law in the 2D-case
\begin{equation}
 \oint_\gamma \ d\zeta_\rho \oint_\gamma \ d\zeta'_{\rho'} \ { D^{\rho\rho'} (\zeta - \zeta')  } 
 = S_\gamma , 
 \label{eq:MM_2}
\end{equation}
\begin{equation}
  \langle 0 |{ {\cal W}_\gamma } | 0 \rangle  
 = {\exp} \left[ {-\frac{g^2}{2} S_\gamma} \right] .  
 \label{eq:MM_3}
\end{equation}
Surely, Eqs. (\ref{eq:MM_1},\ref{eq:MM_2},\ref{eq:MM_3}) are not valid for the `hadronic' Wilson loops (\ref{eq:WL_A}) even in the Abelian case.

To conclude, we have discussed an approach to definition of the generic gluon TMD, which allows us to work directly with the entirely gauge-invariant and maximally path-dependent hadronic matrix elements {\it before} introducing a realistic gluon TMD to be built in an appropriate factorisation framework. All necessary information is absorbed in the basic hadronic correlator (\ref{eq:bas_corr}) and the evolution equation (\ref{eq:SM_MM}) generalises the Makeenko-Migdal equation for the Abelian gauge theory to the case of the `hadronic' loop space. 
Further progress will be reported in a separate work \cite{ChMT_2016}.

\section{Acknowledgements}
The work is supported by the Belgian Federal Science Policy Office. Fruitful discussions of various aspects of this work with T. Mertens, N.G. Stefanis, P. Taels, O.V. Teryaev and F. Van der Veken are gratefully appreciated.


\begin{thebibliography}{99}

\bibitem{Feynman:1969ej}
  R.~P.~Feynman,
  Phys.\ Rev.\ Lett.\  {\bf 23} (1969) 1415; 
 %
  J.~D.~Bjorken and E.~A.~Paschos,
  Phys.\ Rev.\  {\bf 185} (1969) 1975; 
 %
  G.~Curci, W.~Furmanski and R.~Petronzio,
  Nucl.\ Phys.\ B {\bf 175} (1980) 27; 
  %
  Y.~L.~Dokshitzer, D.~Diakonov and S.~I.~Troian,
  Phys.\ Rept.\  {\bf 58} (1980) 269
  

\bibitem{Collins:1981uk}
  J.~C.~Collins and D.~E.~Soper,
  Nucl.\ Phys.\ B {\bf 193} (1981) 381;
   [Nucl.\ Phys.\ B {\bf 213} (1983) 545]

\bibitem{Collins:1981uw}
  J.~C.~Collins and D.~E.~Soper,
  Nucl.\ Phys.\ B {\bf 194} (1982) 445
  
  
 \bibitem{Collins:2011zzd}
  J.~Collins,
  {\it ``Foundations of Perturbative QCD,''}
  Cambridge University Press (2011)
  
 \bibitem{Collins:2003fm}
  J.~C.~Collins,
  Acta Phys.\ Polon.\ B {\bf 34} (2003) 3103
  

  
 \bibitem{Ji:2004wu}
  X.~d.~Ji, J.~p.~Ma and F.~Yuan,
  Phys.\ Rev.\ D {\bf 71} (2005) 034005
  
 \bibitem{Cherednikov:2007tw}
  I.~O.~Cherednikov and N.~G.~Stefanis,
  Phys.\ Rev.\ D {\bf 77} (2008) 094001; 
  Nucl.\ Phys.\ B {\bf 802} (2008) 146;
  %
  Phys.\ Rev.\ D {\bf 80} (2009) 054008;
  %
  I.~O.~Cherednikov, A.~I.~Karanikas and N.~G.~Stefanis,
  Nucl.\ Phys.\ B {\bf 840} (2010) 379

  
 

\bibitem{Boer:2011fh}
  D.~Boer {\it et al.},
  arXiv:1108.1713 [nucl-th]

\bibitem{Angeles-Martinez:2015sea}
  R.~Angeles-Martinez {\it et al.},
  arXiv:1507.05267 [hep-ph]
  
  
\bibitem{Bomhof:2006dp}
  C.~J.~Bomhof, P.~J.~Mulders and F.~Pijlman,
  Eur.\ Phys.\ J.\ C {\bf 47} (2006) 147
  
  \bibitem{Bomhof:2007xt}
  C.~J.~Bomhof and P.~J.~Mulders,
  Nucl.\ Phys.\ B {\bf 795} (2008) 409


\bibitem{Mulders:2000sh}
  P.~J.~Mulders and J.~Rodrigues,
  Phys.\ Rev.\ D {\bf 63} (2001) 094021
  
    \bibitem{Belitsky:2002sm}
  A.~V.~Belitsky, X.~Ji and F.~Yuan,
  Nucl.\ Phys.\ B {\bf 656} (2003) 165;
%
  D.~Boer, P.~J.~Mulders and F.~Pijlman,
  Nucl.\ Phys.\ B {\bf 667} (2003) 201;
  A.~V.~Belitsky and A.~V.~Radyushkin,
  Phys.\ Rept.\  {\bf 418} (2005) 1
  %

  
  \bibitem{Ji:2005nu}
  X.~d.~Ji, J.~P.~Ma and F.~Yuan,
  JHEP {\bf 0507} (2005) 020
  

\bibitem{Dominguez:2011wm}
  F.~Dominguez, C.~Marquet, B.~W.~Xiao and F.~Yuan,
  Phys.\ Rev.\ D {\bf 83} (2011) 105005
  
 \bibitem{Balitsky:2015qba}
  I.~Balitsky and A.~Tarasov,
  JHEP {\bf 1510} (2015) 017; 
  Int.\ J.\ Mod.\ Phys.\ Conf.\ Ser.\  {\bf 37} (2015) 1560058;
  I.~Balitsky,
  arXiv:1510.06430 [hep-ph]



    \bibitem{MM_WL_1}
Y.~M.~Makeenko and A.~A.~Migdal,
  Phys.\ Lett.\ B {\bf 88} (1979) 135, 
   [Erratum-ibid.\ B {\bf 89} (1980) 437]; 
%
  Nucl.\ Phys.\ B {\bf 188} (1981) 269
  
  \bibitem{Makeenko:2002uj}
  Y.~Makeenko,
 {\it ``Methods of Contemporary Gauge Theory''},  
 Cambridge University Press (2002)



  \bibitem{WL_Renorm_1}
 R.~A.~Brandt, F.~Neri and M.~A.~Sato,
  Phys.\ Rev.\ D {\bf 24} (1981) 879;
%
 R.~A.~Brandt, A.~Gocksch, M.~A.~Sato and F.~Neri,
  Phys.\ Rev.\ D {\bf 26} (1982) 3611

\bibitem{Tavares:1993pw}
  J.~N.~Tavares,
  Int.\ J.\ Mod.\ Phys.\ A {\bf 9} (1994) 4511

\bibitem{DeG_2014}
I.O. Cherednikov, T. Mertens and F.F. Van der Veken, {\it ``Wilson Lines in Quantum Field Theory''},
De Gruyter, Berlin (2014)



\bibitem{Cherednikov:2012yd}
  I.~O.~Cherednikov, T.~Mertens and F.~F.~Van der Veken,
  Phys.\ Rev.\ D {\bf 86} (2012) 085035;
%
%
  Phys.\ Part.\ Nucl.\  {\bf 44} (2013) 250

  \bibitem{Mertens:2013xga}
  T.~Mertens and P.~Taels,
  Phys.\ Lett.\ B {\bf 727} (2013) 563; 
  I.~Cherednikov and T.~Mertens,
  Phys.\ Lett.\ B {\bf 734} (2014) 198; 
  Phys.\ Lett.\ B {\bf 741} (2015) 71;
  I.~O.~Cherednikov,
  Few Body Syst.\  {\bf 55} (2014) 303
 
 \bibitem{Gatheral:1983cz}
  J.~G.~M.~Gatheral,
  Phys.\ Lett.\ B {\bf 133} (1983) 90; 
%
  J.~Frenkel and J.~C.~Taylor,
  Nucl.\ Phys.\ B {\bf 246} (1984) 231; 
 %
  G.~P.~Korchemsky and A.~V.~Radyushkin,
  Nucl.\ Phys.\ B {\bf 283} (1987) 342;
  E.~Laenen, G.~Stavenga and C.~D.~White,
  JHEP {\bf 0903} (2009) 054;
 %
  A.~Mitov, G.~Sterman and I.~Sung,
  Phys.\ Rev.\ D {\bf 82} (2010) 096010;
  %
  E.~Gardi, E.~Laenen, G.~Stavenga and C.~D.~White,
  JHEP {\bf 1011} (2010) 155; 
 %
  A.~A.~Vladimirov,
  JHEP {\bf 1506} (2015) 120
 
 
\bibitem{ChMT_2016} I.O. Cherednikov, T. Mertens and P. Taels, {\it [in preparation]}

 

\end{thebibliography}
\end{document}